\documentclass{vgtc}                          

\graphicspath{{figures/}{pictures/}{images/}{./}} 

\usepackage{times}                     

\usepackage{tabu}                      
\usepackage{booktabs}                  
\usepackage{lipsum}                    
\usepackage{mwe}                       

\usepackage{subcaption}

\usepackage{mathptmx}                  

\onlineid{0}

\vgtccategory{Research}

\vgtcinsertpkg

\title{Beyond Task Performance: Lessons Learned from Evaluating an Exploratory VR Interaction Technique}

\author{
Nevzat Umut Demirseren\thanks{e-mail: ndemirseren@txstate.edu}\\
\scriptsize Texas State University
\and
Corey Pittman\thanks{e-mail: corey.pittman@ucf.edu}\\
\scriptsize University of Central Florida
\and
Isayas Berhe Adhanom\thanks{e-mail: isayas@txstate.edu}\\
\scriptsize Texas State University
\and
Karthikeyan Umapathy\thanks{e-mail: k.umapathy@unf.edu}\\
\scriptsize University of North Florida
\and
Kevin Pfeil\thanks{e-mail: kevin.pfeil@ucf.edu}\\
\scriptsize University of Central Florida
}

\abstract{
Dense virtual environments present significant challenges for object selection and manipulation, motivating the development of novel interaction techniques. This paper presents RodCast as an exploratory case study to investigate explicit trajectory visualization for interaction in dense virtual environments. We conducted a within-subjects user study comparing RodCast with Go-Go Hand and FlowerCone across three representative interaction tasks using objective performance measures, subjective evaluations, and qualitative feedback. The results revealed that the proposed implementation incurred performance costs on more demanding manipulation tasks, while qualitative feedback suggested that participants attributed these challenges to the complexity of the bimanual control scheme. Additionally, subjective evaluations and qualitative feedback identified benefits in spatial awareness and target accessibility that were not fully reflected by conventional performance measures. Together, these findings highlight the importance of control simplicity in interaction technique design and emphasize that incorporating qualitative feedback into the evaluation process is essential for distinguishing implementation limitations from the potential of the interaction concept. Collectively, these lessons provide guidance for the design and evaluation of future interaction techniques.

} 

\keywords{Virtual Reality, Interaction Evaluation, Exploratory Interaction Techniques, Spatial Awareness, User Experience}

\begin{document}


\firstsection{Introduction}

\maketitle



Virtual reality (VR) interaction techniques continue to evolve toward supporting increasingly complex tasks in dense virtual environments (VEs). Object selection and manipulation in cluttered scenes remain particularly challenging due to visual occlusion, target ambiguity, and limited accessibility \cite{vanacken2009multimodal, yu2020fully}. Existing interaction techniques primarily seek to improve efficiency and precision under these conditions, yet balancing high task performance with users' understanding of complex spatial relationships remains an important design challenge.

Prior work has shown that visual guidance and interaction design can influence users' performance and experience during VR interaction \cite{abtahi2022beyond, harada2022quantitative}. Motivated by these observations, we explore this design space through \textit{RodCast Interaction}, a bimanual interaction technique that employs an explicit arc-shaped interaction trajectory to access partially or fully occluded targets. Rather than relying solely on direct ray-based interaction, RodCast visualizes the complete interaction path, enabling users to continuously manipulate the trajectory.

We evaluated RodCast against two established interaction techniques through a controlled user study involving distance perception, dense manipulation, and object sorting tasks. The resulting performance tradeoffs were unexpected. Although RodCast did not consistently outperform existing techniques on conventional efficiency and manipulation measures, participants reported improved spatial awareness and understanding of target accessibility, and the technique demonstrated advantages during spatial localization tasks. These findings raise a broader question: \emph{Are conventional evaluation metrics sufficient for assessing interaction techniques whose primary contributions extend beyond task efficiency?}

Rather than presenting RodCast as a definitive replacement for existing interaction techniques, this paper uses it as an exploratory case study to examine the challenges of designing and evaluating interaction concepts. The observed tradeoffs highlight the importance of simple control schemes and demonstrate how incorporating qualitative feedback into the evaluation process can provide a deeper understanding of design tradeoffs and implementation limitations. Together, these lessons provide practical guidance for future interaction design and evaluation.

\section{Background}

\subsection{Evaluation of Interaction Techniques in VR}

Interaction techniques in VR are commonly evaluated using objective performance measures such as completion time, selection accuracy, and error rate, providing standardized comparisons of efficiency and precision across interaction paradigms~\cite{laviola20173d, mackenzie2013human}. While these metrics have been instrumental in advancing 3D user interfaces, interaction techniques often seek to improve aspects of the interaction experience beyond efficiency and precision by providing additional visual guidance and supporting users' interpretation of complex spatial interactions \cite{abtahi2022beyond, harada2022quantitative}. Because spatial interaction fundamentally depends on users' perception of depth, object location, and spatial relationships~\cite{cutting1995perceiving, loomis2003visual, masnadi2022effects}, techniques that provide explicit spatial feedback may offer meaningful design benefits that are not fully reflected by conventional performance measures alone. Consequently, exploratory interaction techniques may contribute valuable design insights by exposing tradeoffs between efficiency, learnability, and coordination effort, motivating broader perspectives on how VR interaction techniques are evaluated.

\subsection{Interaction in Dense VEs}

Object selection in dense VEs remains challenging due to visual occlusion, target ambiguity, and increased cognitive demand, which are further compounded by limitations in depth perception~\cite{cutting1995perceiving, laviola20173d, vanacken2009multimodal}. Conventional ray-casting techniques provide intuitive distant interaction but often struggle to accurately access partially or fully occluded targets~\cite{bowman1997evaluation, kopper2011rapid}. To address these limitations, researchers have proposed volume-based selection methods, two-step refinement techniques, and visual augmentation strategies that improve target disambiguation and accessibility, often at the cost of increased interaction complexity or additional interaction overhead~\cite{forsberg1996aperture, grossman2005bubble, krekhov2018deadeye, wang2025focalselect, yu2020fully}.

More recently, interaction techniques have explored redirected or curved interaction trajectories, such as Folding Rays~\cite{kim2023folding}, EEBA~\cite{wu2024eeba}, and vMirror~\cite{li2021vmirror}, to improve access to occluded targets while providing richer spatial guidance. 
Likewise, bimanual interaction has been shown to support asymmetric coordination through the complementary roles of both hands, while fine motor control and compatible control mappings facilitate precise spatial interaction~\cite{guiard1987asymmetric, zhai1996human}.
Collectively, these approaches illustrate that interaction design for dense environments requires balancing multiple competing objectives, including efficiency, precision, accessibility, coordination effort, and users' understanding of the interaction space. These tradeoffs motivate the exploration of alternative interaction paradigms and raise broader questions about how novel interaction techniques should be evaluated when their primary contributions extend beyond conventional task-performance metrics.

\section{Methodology}

\subsection{RodCast Interaction}

RodCast Interaction\footnote{\url{github.com/TheCrimsonNeV0/RodCast_Lab}} is an exploratory VR interaction technique designed to investigate the tradeoffs of combining explicit arc-guided interaction with bimanual control for object selection and manipulation in dense VEs. Inspired by the Flexible Pointer~~\cite{feiner2003flexible}, which enables access to partially occluded targets through a deformable curved ray, RodCast replaces explicit curve-point manipulation with continuous bimanual control of an arc-shaped interaction trajectory. The visible trajectory and supplementary depth cues were intended to improve users' understanding of target accessibility and spatial relationships, while serving as a case study for examining the benefits and challenges of arc-guided interaction.

\begin{figure}[htbp]
    \centering
    \begin{subfigure}[b]{0.493\columnwidth}
        \centering
        \includegraphics[width=\linewidth]{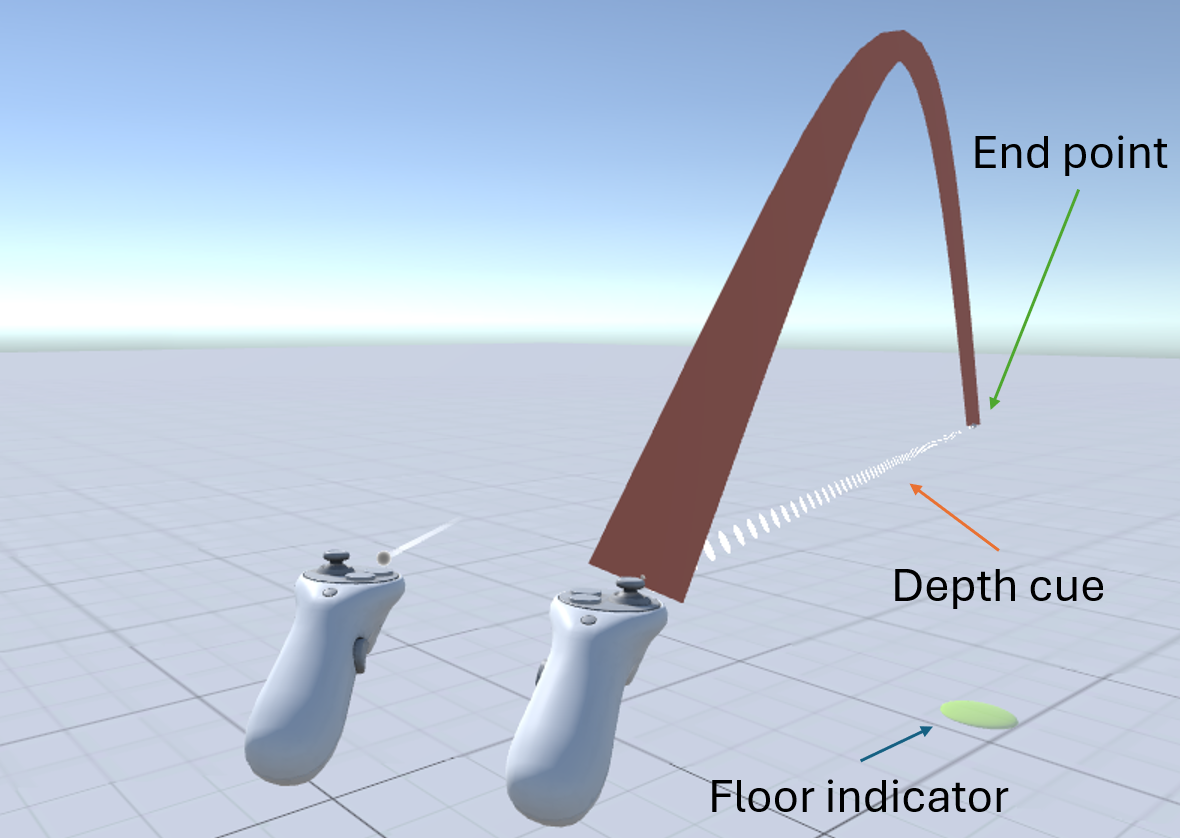}
        \caption{Arc-guided pointer.}
        \label{fig:rodcast_arc}
    \end{subfigure}
    \hfill
    \begin{subfigure}[b]{0.493\columnwidth}
        \centering
        \includegraphics[width=\linewidth]{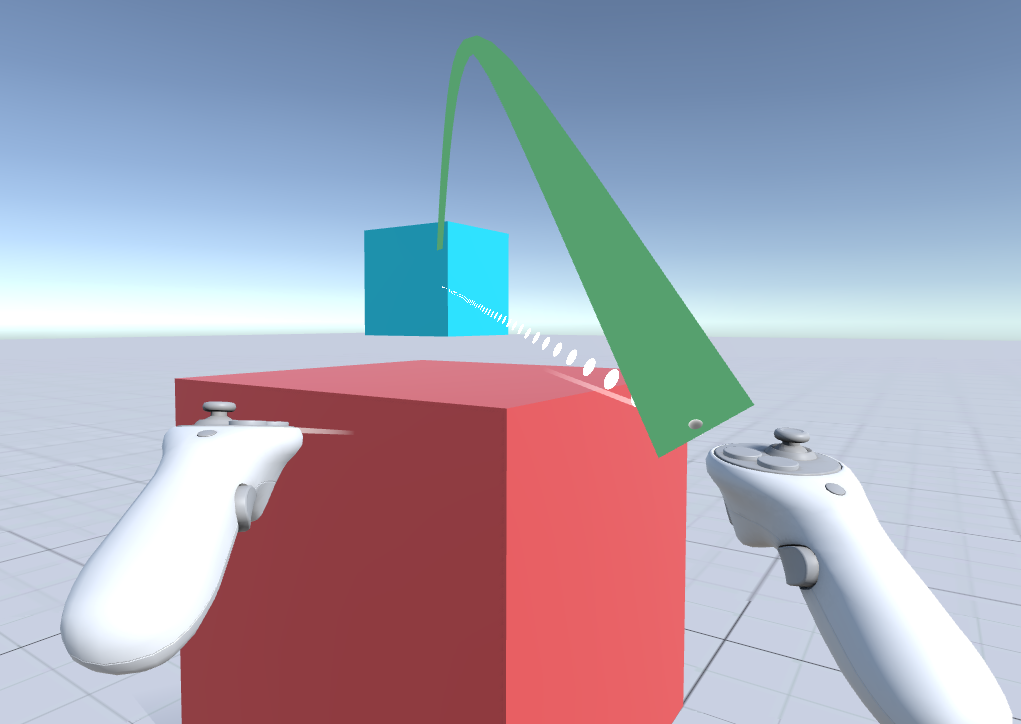}
        \caption{Object manipulation.}
        \label{fig:rodcast_object}
    \end{subfigure}
    \caption{RodCast technique (left) and object manipulation (right).}
    \label{fig:rodcast}
\end{figure}

\subsubsection{Interaction Design}

RodCast was designed to explore how explicit trajectory visualization influences users' understanding of spatial relationships during interaction in dense VEs. Rather than relying on a conventional straight ray, RodCast employs a visible arc-shaped interaction trajectory that users continuously manipulate through coordinated bimanual gestures. The arc provides an explicit representation of the interaction path, communicating target accessibility, reachability, and the spatial consequences of users' actions while supporting indirect access to partially or fully occluded objects. Supplementary visual cues, including a floor projection indicator and dynamic interaction feedback, further reinforce depth perception and spatial awareness throughout object manipulation. By combining arc-guided interaction with coordinated bimanual control, RodCast was intended to investigate whether explicit spatial guidance can enhance users' understanding of complex interaction spaces, even if doing so introduces additional coordination effort.


\subsubsection{Control Scheme}

RodCast employs an asymmetric bimanual control scheme in which the dominant hand is responsible for interaction and endpoint positioning, while the non-dominant hand continuously shapes the interaction trajectory. The dominant hand adjusts the reachable endpoint of the arc, initiates object selection, and retrieves objects along the visible trajectory. Selection is only permitted when the arc reaches the target through an unobstructed path, reinforcing the relationship between the visualized trajectory and the resulting interaction.

The non-dominant hand controls the overall shape of the arc by adjusting its length, curvature, and orientation through continuous spatial gestures, enabling users to redirect the interaction path around occluding objects. Once an object is selected, coordinated bimanual manipulation allows users to reposition both the trajectory and the object simultaneously. This asymmetric division of control was intended to separate trajectory planning from object manipulation, encouraging users to explicitly reason about reachability and spatial relationships while interacting within cluttered environments.

\subsection{Study Design}

We conducted a $3 \times 3$ within-subjects user study comparing three VR interaction techniques across three representative interaction tasks, with each task including additional within-subject experimental conditions. Participant sample size was determined using \textit{a priori} power analysis. 22 participants (18 male, 4 female; mean age = 26.1 years), all with normal or corrected-to-normal vision, completed every combination of technique and task in randomized order. Participants reported moderate familiarity with 3D video games ($M = 3.27/5$) but relatively limited prior VR experience ($M = 2.0/5$). The study was conducted in a dedicated VR research laboratory using a Meta Quest 3 headset (2064 × 2208 pixels per eye, 90 Hz) with individualized IPD calibration. Participants remained stationary throughout the study and were permitted to complete the tasks while either seated or standing. Each task included task-specific experimental variations, with target distance manipulated in the distance perception task and object density varied in the dense manipulation and object sorting tasks. Objective performance measures and post-study usability reports were collected to characterize the strengths and tradeoffs of each interaction technique. The study was approved by the University of North Florida Institutional Review Board (IRB), and all participants provided informed consent prior to participation.

\begin{figure*}[t]
    \centering

    \begin{subfigure}[t]{0.327\textwidth}
        \centering
        \includegraphics[width=\linewidth]{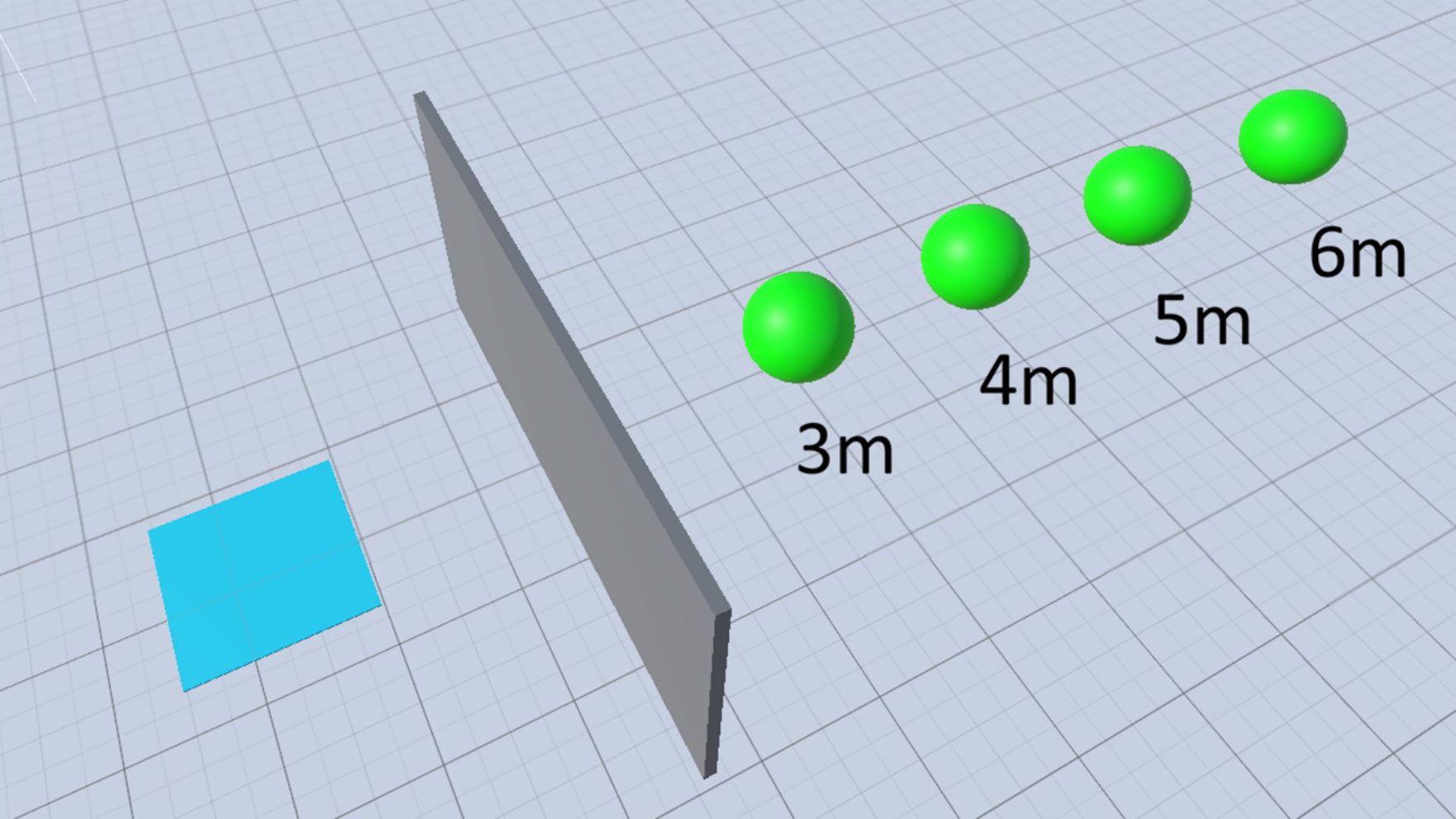}
        \caption{Distance Perception Task}
        \label{fig:distance_task}
    \end{subfigure}
    \hfill
    \begin{subfigure}[t]{0.327\textwidth}
        \centering
        \includegraphics[width=\linewidth]{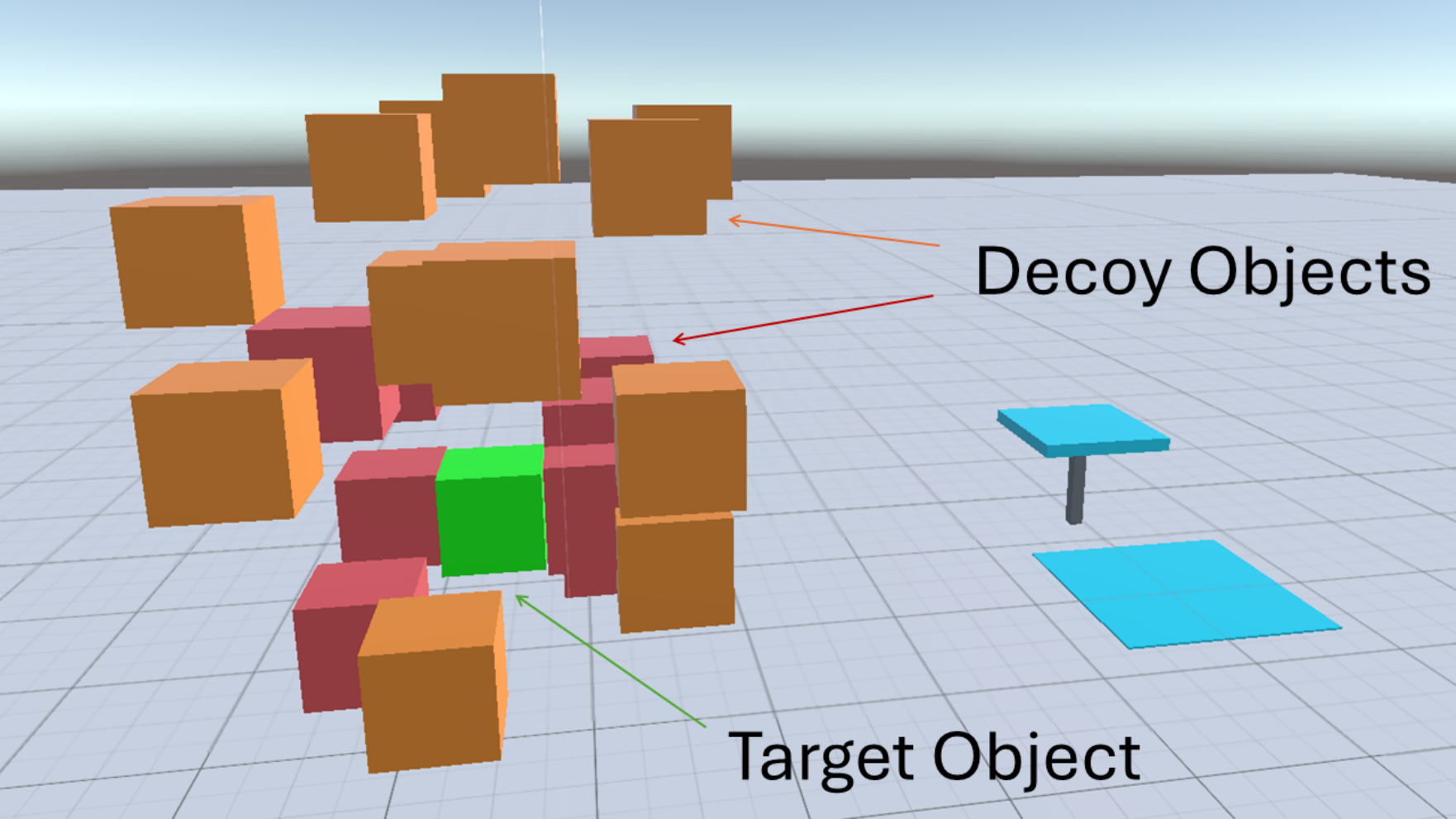}
        \caption{Dense Manipulation Task}
        \label{fig:dense_task}
    \end{subfigure}
    \hfill
    \begin{subfigure}[t]{0.327\textwidth}
        \centering
        \includegraphics[width=\linewidth]{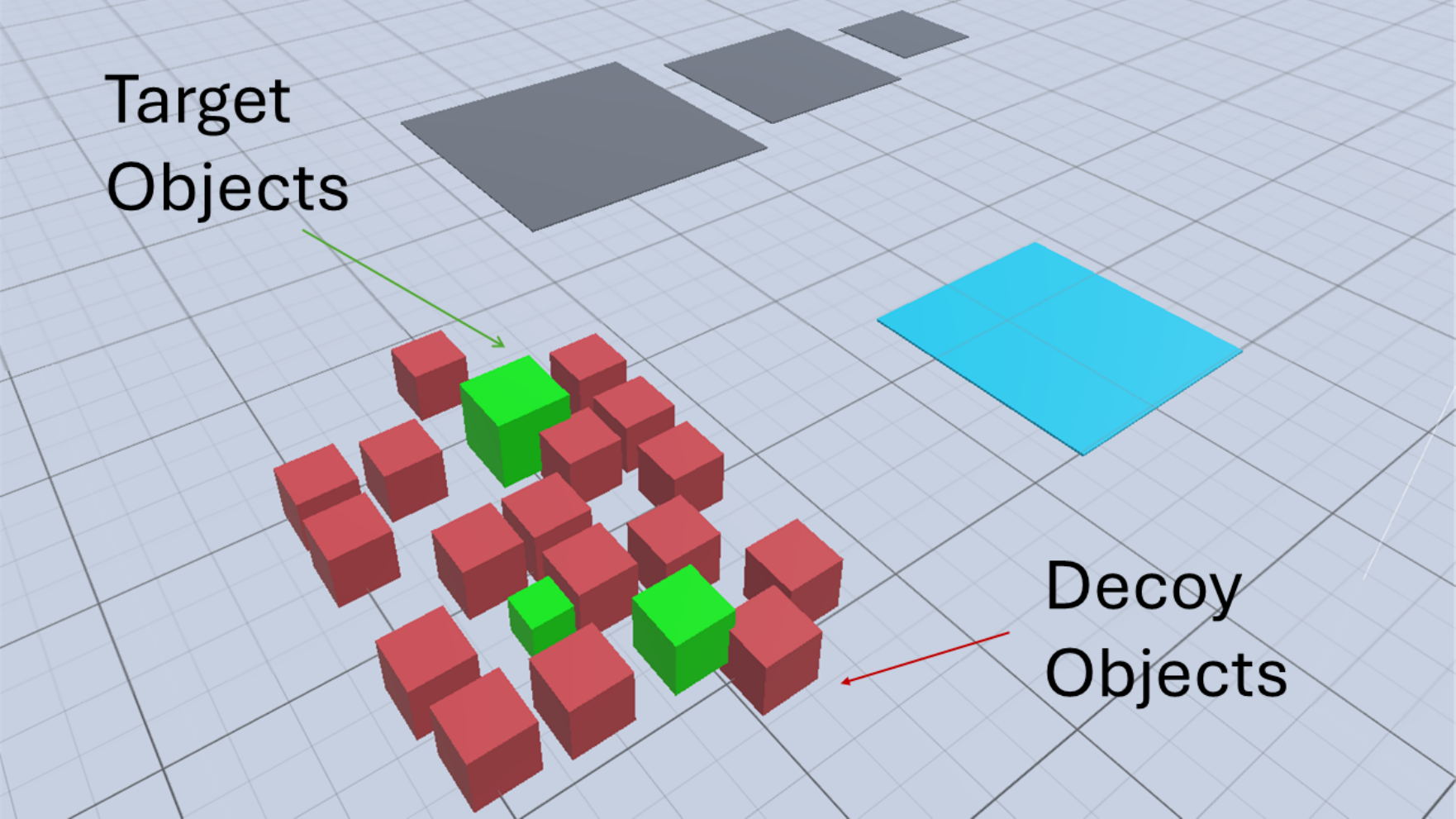}
        \caption{Object Sorting Task}
        \label{fig:sorting_task}
    \end{subfigure}

    \caption{Virtual environments used for the three experimental tasks: (a) Distance Perception, (b) Dense Manipulation, and (c) Object Sorting.}
    \label{fig:task_environments}
\end{figure*}

\subsubsection{Compared Techniques}

RodCast was evaluated against Go-Go Hand~\cite{poupyrev1996go} and FlowerCone~\cite{yu2020fully} as representative implementations of the widely used reach-extension and volume-based interaction paradigms for distant and occluded object interaction in VR. Go-Go Hand served as a reach-extension baseline, while FlowerCone represented a volume-based selection technique designed for dense environments through cone-based target acquisition and refinement.

To better align FlowerCone with the continuous manipulation tasks used in our study, we adapted its original implementation by replacing the two-step refinement phase with automatic highlighting of the object nearest the center of the selection volume,while allowing users to cycle through additional candidate targets via a dedicated controller button. We also incorporated a floor projection indicator, similar to RodCast, to provide comparable depth feedback during interaction. These modifications were intended to preserve FlowerCone's core volume-based selection behavior while reducing procedural differences that were not central to the interaction concepts being compared. The modified implementation was pilot tested to verify that its overall interaction characteristics remained consistent with the original technique.

\subsubsection{User Tasks}

Participants completed three representative interaction tasks designed to capture complementary challenges commonly encountered in dense VEs. The tasks examined spatial localization under occlusion, precise object manipulation within clutter, and repeated target acquisition during extended interaction. All conditions were repeated three times and presented in randomized, non-consecutive order to reduce potential ordering effects~\cite{lazar2017research, mackenzie2013human}. A five-second countdown preceded each trial.

\begin{enumerate}
    \item \textbf{Distance Perception:} Participants observed the target before it became fully occluded. After the target was occluded, the assigned interaction technique was revealed, with all technique-specific supplementary visual cues remained disabled throughout the task. Using the assigned interaction technique, participants estimated the remembered target location. Target distances of 3, 4, 5, and 6~m were selected following the range of distances used in prior work~\cite{hmaiti2024visual}.


    \item \textbf{Dense Manipulation:} 
    Participants retrieved a target object randomly positioned within a highlighted region and transported it to the target surface while avoiding unintended collisions with surrounding immovable decoy objects. Three density conditions progressively increased visual clutter and target occlusion: low density (4--7 inner decoys), medium density (4--6 inner and 10--15 outer decoys), and high density (5--7 inner and 10--15 outer decoys).


    \item \textbf{Object Sorting:} Participants repeatedly located three partially or fully occluded target objects of varying sizes and sorted them onto their corresponding destinations as efficiently as possible. Unlike the dense manipulation task, all objects were movable in this task. Three density conditions progressively increased clutter: low density (4--7 inner objects; 8--12 total objects), medium density (4--6 inner and 10--15 outer objects; 15--19 total objects), and high density (5--7 inner and 10--15 outer objects; 20--25 total objects).


    
\end{enumerate}

\subsubsection{Measures}

The evaluation combined performance measures with subjective usability assessments to provide complementary perspectives on each interaction technique. Objective measures quantified interaction performance, while subjective assessments captured participants' perceived usability.

\begin{itemize}
    \item \textbf{Distance Estimation Offset (Distance Perception):} Euclidean distance (m) between the target location and the participant's estimated position, measuring spatial localization accuracy under full occlusion.

    \item \textbf{Task Completion Time (Dense Manipulation \& Object Sorting):} Time (s) required to successfully complete each task, measuring interaction efficiency.

    \item \textbf{Decoy Touch Count (Dense Manipulation):} Number of unintended collisions between the manipulated target and surrounding distractor objects, measuring interaction precision in cluttered environments.

    \item \textbf{Subjective Usability (Post-study):} A post-study questionnaire assessing perceived usability, learnability, spatial feedback, and overall interaction experience across all techniques.
\end{itemize}

\section{Results}

\subsection{Objective Performance}

Incomplete trials and trials with data loss were excluded prior to analysis. Shapiro--Wilk test \cite{shapiro1965analysis} results indicated non-normality for all pairwise comparisons; therefore, Wilcoxon signed-rank tests \cite{wilcoxon1992individual} were used for all post hoc analyses.

\subsubsection{Distance Perception}

A repeated-measures ANOVA revealed significant effects of interaction technique ($F(2,42)=16.88$, $p<.001$, $\eta_p^2=.20$), target distance ($F(3,63)=10.41$, $p<.001$, $\eta_p^2=.05$), and their interaction ($F(6,126)=3.02$, $p=.009$, $\eta_p^2=.036$) on distance estimation offset. Post-hoc pairwise comparisons using Wilcoxon signed-rank tests with Holm correction showed that both RodCast ($Z=-5.89$, $p<.001$, $d_z=0.62$) and Go-Go Hand ($Z=-6.70$, $p<.001$, $d_z=0.70$) produced significantly lower localization error than FlowerCone. No significant difference was observed between RodCast and Go-Go Hand ($Z=-1.79$, $p=.074$, $d_z=0.24$).

\begin{figure}[h]
  \centering
  \includegraphics[width=1.00\linewidth]{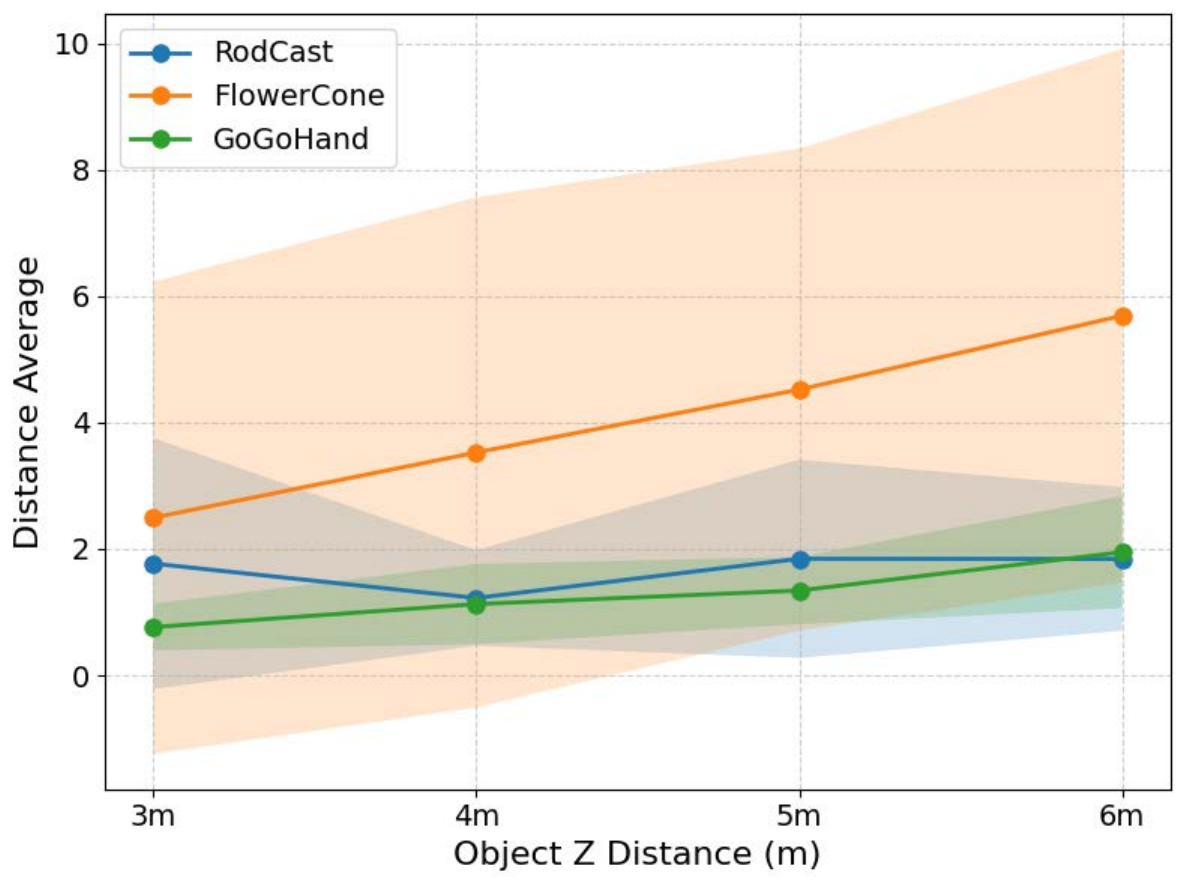}
\caption{Mean distance estimation offset for the Distance Perception task across interaction techniques and target distances. Shaded regions represent $\pm1$ standard deviation.}
  \label{fig:environment}
\end{figure}

\subsubsection{Dense Manipulation}

Separate repeated-measures ANOVAs revealed significant effects of interaction technique on both task completion time ($F(2,42)=8.44$, $p=.001$, $\eta_p^2=.038$) and decoy touch count ($F(2,42)=5.79$, $p=.006$, $\eta_p^2=.042$). Density level significantly affected decoy touch count ($F(2,42)=7.79$, $p=.002$, $\eta_p^2=.050$), but not task completion time ($F(2,42)=1.65$, $p=.204$, $\eta_p^2=.010$), and no significant interaction effects between technique and density were observed for either task completion time ($F(4,84)=0.47$, $p=.756$, $\eta_p^2=.005$) or decoy touch count ($F(4,84)=0.68$, $p=.520$, $\eta_p^2=.015$). Post-hoc pairwise comparisons using Wilcoxon signed-rank tests with Holm correction showed that RodCast produced significantly longer completion times and higher decoy touch counts than both FlowerCone (completion time: $Z=-3.42$, $p=.002$, $d_z=0.47$; decoy touch count: $Z=-2.86$, $p=.008$, $d_z=0.30$) and Go-Go Hand (completion time: $Z=-2.83$, $p=.009$, $d_z=0.32$; decoy touch count: $Z=-4.09$, $p<.001$, $d_z=0.34$). No significant differences were observed between FlowerCone and Go-Go Hand for either measure (completion time: $Z=-0.53$, $p=.594$, $d_z=0.13$; decoy touch count: $Z=-0.12$, $p=.904$, $d_z=0.03$).

\begin{figure}[htbp]
    \centering
    \includegraphics[width=0.72\linewidth]{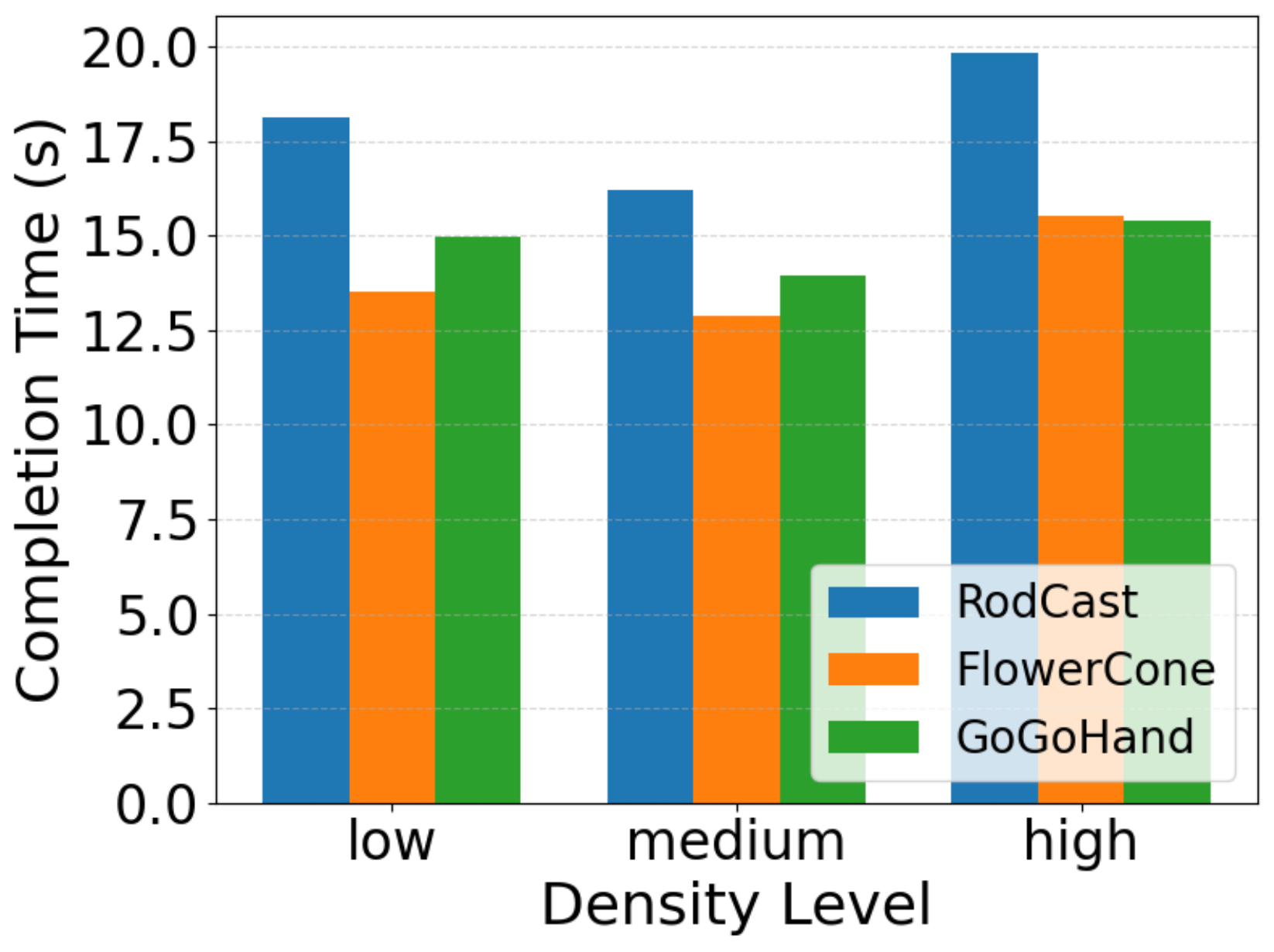}
    \caption{Mean decoy touch count during the Dense Manipulation task for each interaction technique across density levels.}
    \label{fig:densemanip_touches}
\end{figure}

\begin{figure}[htbp]
    \centering
    \includegraphics[width=0.70\linewidth]{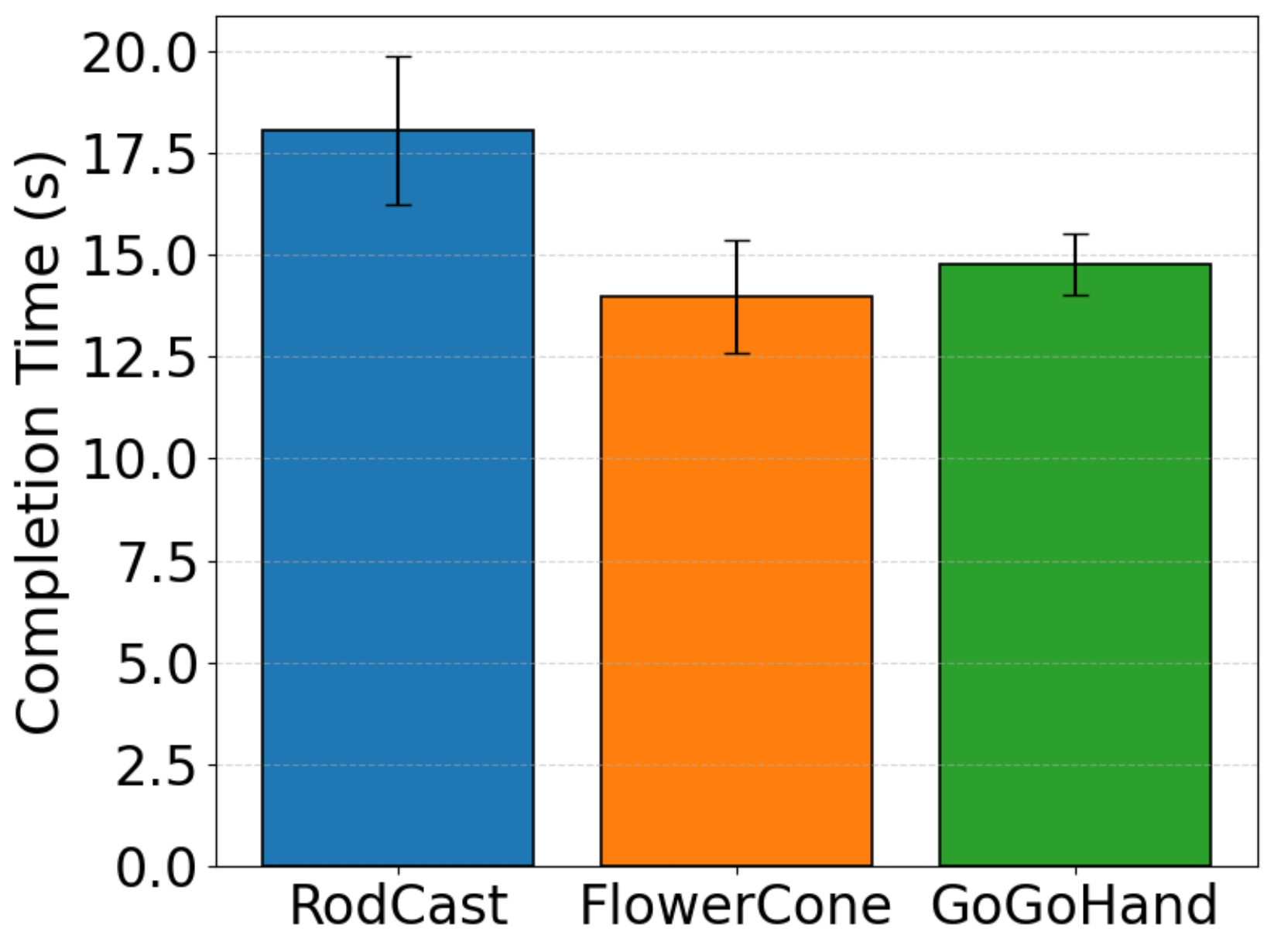}
    \caption{Mean task completion time for each interaction technique across all object density levels during the Dense Manipulation task. Error bars represent standard errors.}
    \label{fig:densemanip_time}
\end{figure}

\subsubsection{Object Sorting}

A repeated-measures ANOVA revealed a significant effect of interaction technique on task completion time ($F(2,42)=10.56$, $p<.001$, $\eta_p^2=.335$). Post-hoc pairwise comparisons using Wilcoxon signed-rank tests with Holm correction showed that FlowerCone produced significantly shorter completion times than both RodCast ($Z=3.93$, $p<.001$, $d_z=0.84$) and Go-Go Hand ($Z=-3.67$, $p<.001$, $d_z=0.81$). No significant difference was observed between RodCast and Go-Go Hand ($Z=0.55$, $p=.581$, $d_z=0.08$).

\begin{figure}[h]
  \centering
  \includegraphics[width=0.70\linewidth]{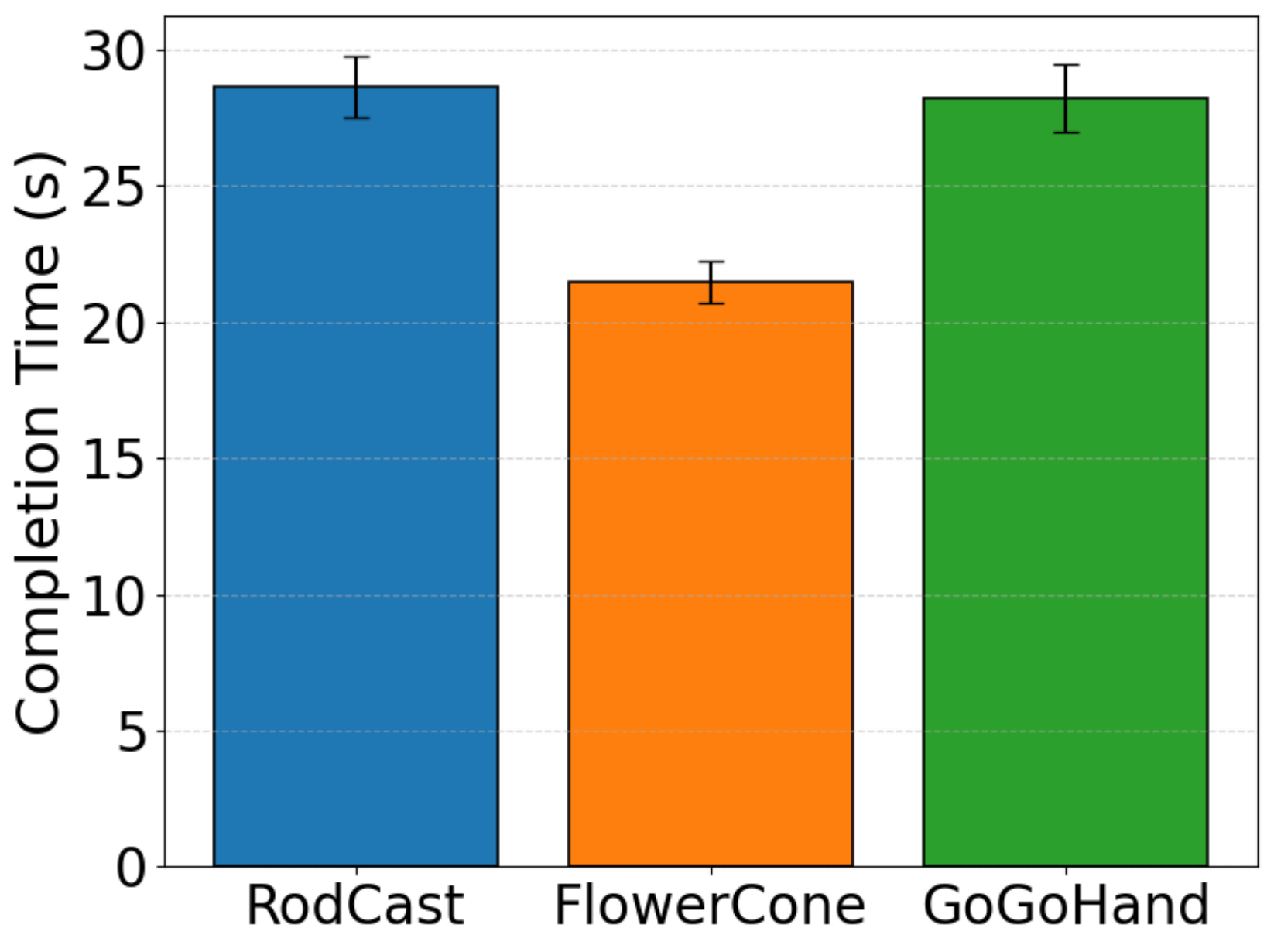}
\caption{Mean task completion time for each interaction technique during the Object Sorting task. Error bars represent standard errors.}
  \label{fig:environment}
\end{figure}

\subsection{Subjective Usability}

\subsubsection{Post-Study Preference Rankings}


Participants ranked the interaction techniques across eight usability dimensions, where lower rankings indicate higher preference. Go-Go Hand received the lowest mean rankings for ease of use (1.55), comfort and ergonomics (1.68), precision and accuracy (1.77), speed of interaction (1.73), learnability (1.45), immersion (1.36), and user satisfaction (1.59), whereas FlowerCone received the lowest mean ranking for feedback and responsiveness (1.77). Friedman tests identified significant overall differences for ease of use ($\chi^2=11.27$, $p=.004$), comfort and ergonomics ($\chi^2=6.09$, $p=.048$), immersion ($\chi^2=13.36$, $p=.001$), learnability ($\chi^2=10.64$, $p=.005$), and user satisfaction ($\chi^2=6.64$, $p=.036$), whereas no significant differences were observed for speed of interaction ($p=.195$), precision and accuracy ($p=.351$), or feedback and responsiveness ($p=.422$). Post-hoc Wilcoxon signed-rank tests with Holm correction revealed significant differences only between RodCast and Go-Go Hand for ease of use ($p=.004$), immersion ($p=.006$), and learnability ($p=.021$). No statistically significant differences were observed between RodCast and FlowerCone for any usability dimension.

\begin{figure}[h]
  \centering
  \includegraphics[width=0.98\linewidth]{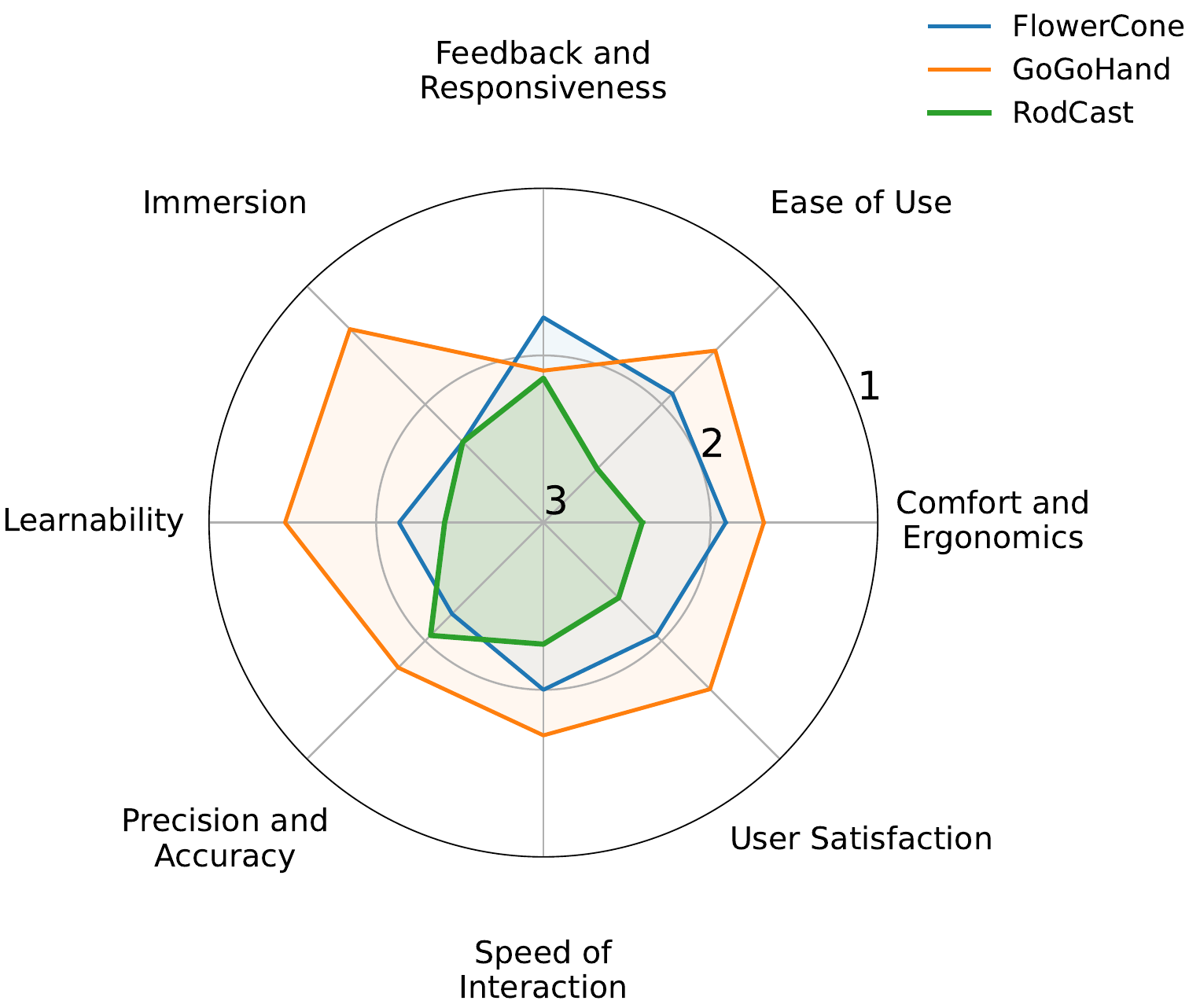}
  \caption{Radar plot comparing the mean preference rankings of each interaction technique across subjective usability dimensions.}
  \label{fig:environment}
\end{figure}

\subsubsection{Qualitative Feedback}

Open-ended feedback provided additional context for interpreting the quantitative findings and highlighted design tradeoffs not fully captured by the objective measures. Many participants reported that the visible arc trajectory and supplementary depth cues improved awareness of target accessibility and spatial relationships in cluttered environments, supporting reasoning about interaction paths beyond direct line-of-sight. However, participants consistently identified the bimanual control scheme as the primary usability challenge, noting that coordinating both hands and trigger mappings required additional effort and practice. More experienced VR users often viewed this coordination as beneficial for precise manipulation, whereas less experienced users more frequently reported a steep learning curve. Several participants also suggested alternative control mappings, such as assigning the interaction ray to the non-dominant hand, and one participant with extensive prior VR experience who adopted this reversed mapping reported greater comfort than with the intended configuration. Collectively, these observations suggest that the primary opportunity for refinement lies in simplifying the control mapping rather than the underlying arc-guided interaction concept itself.

\section{Discussion}

The objective evaluation presents a nuanced view of RodCast's design. While the technique did not outperform FlowerCone in dense manipulation or repeated object sorting, it achieved significantly lower localization error than FlowerCone during the distance perception task. This suggests that the trajectory visualization can support spatial reasoning even when it does not maximize manipulation efficiency. Rather than demonstrating a uniformly superior interaction technique, these findings reveal that different aspects of interaction benefits from different design priorities. Conventional performance measures captured the additional coordination overhead introduced by RodCast's bimanual control scheme, whereas the distance perception results and participant feedback indicate that the visible arc provided meaningful support for understanding object reachability and spatial relationships in cluttered environments. These findings suggest that introducing explicit trajectory guidance can improve users' spatial understanding, but that these benefits depend critically on how the interaction is mapped to user input.

The qualitative feedback further clarified the source of this tradeoff. Participants consistently attributed the additional interaction difficulty to the control mapping rather than to the curved trajectory itself. In particular, many inexperienced VR users reported that coordinating two triggers while simultaneously controlling the arc shape introduced unnecessary cognitive load, making the technique more difficult to learn during the limited study period. Conversely, participants with greater VR experience adapted more readily to the interaction and frequently described the arc visualization as beneficial for precise positioning, target accessibility, and spatial awareness. Together, these observations suggest that the primary design challenge was not introducing explicit trajectory guidance, but integrating it into a control scheme that minimizes coordination demands while preserving its spatial benefits.

The principal lesson from this study extends beyond RodCast itself. Our findings suggest that the interaction concept and its input mapping should be treated as separate aspects of VR interaction design.
While the current dual-trigger bimanual control scheme introduced measurable performance costs, particularly during complex manipulation tasks and for users with limited prior VR experience, the objective, subjective, and qualitative findings collectively indicate that the underlying concept of explicit trajectory visualization remains promising. RodCast demonstrated improved distance localization, while participants consistently described the visible arc as improving their understanding of target accessibility, spatial relationships, and perceived interaction precision. Future work should therefore explore alternative control mappings that preserve explicit trajectory guidance while reducing coordination complexity.

More broadly, this work raises important questions about how exploratory VR interaction techniques should be evaluated. Conventional performance metrics such as task completion time and manipulation accuracy remain essential, but they may not fully capture the potential value of unfamiliar interaction paradigms during initial exposure, particularly when substantial learning or adaptation is required before their intended benefits emerge. In such cases, subjective assessments, qualitative feedback, and evaluations targeting the specific design objectives of a technique can provide complementary evidence that helps distinguish limitations of an implementation from the promise of the underlying interaction concept. From this perspective, RodCast should not be viewed as a failed interaction technique, but as an exploration of a design space that highlights both the challenges of implementing explicit trajectory guidance through bimanual interaction and the importance of evaluating experimental VR interaction techniques beyond immediate performance outcomes.

\section{Conclusion}

This paper presented RodCast as an exploratory case study of arc-guided interaction for dense virtual environments. The evaluation revealed that while RodCast improved distance perception and supported effective target selection through explicit trajectory visualization, the complexity of its dual-trigger bimanual control scheme introduced a steep learning curve that limited performance during more demanding manipulation tasks, particularly for users with limited prior VR experience. At the same time, participants consistently associated the visible arc with improved spatial awareness and target accessibility, suggesting that some advantages of the interaction concept were not fully reflected by conventional task-level performance measures. While acknowledging the usability limitations of the current implementation, consistent positive feedback, particularly from participants with greater prior VR experience, raises a broader question about how effectively conventional objective metrics capture the full range of benefits offered by experimental VR interaction techniques. We hope this work encourages broader discussion of both bimanual interaction design and evaluation methodologies that combine objective performance with subjective experience, qualitative feedback, and assessments targeting the intended interaction goals, enabling researchers to better distinguish the potential of an interaction concept from limitations of its implementation.


\bibliographystyle{abbrv-doi}

\bibliography{template}
\end{document}